\documentstyle[floats,aps]{revtex}
\begin{document}
\input epsf

\draft

  \font\twelvemib=cmmib10 scaled 1200
  \font\elevenmib=cmmib10 scaled 1095
  \font\tenmib=cmmib10
  \font\eightmib=cmmib10 scaled 800
  \font\sixmib=cmmib10 scaled 667
  \skewchar\elevenmib='177
  \newfam\mibfam
  \def\mib{\fam\mibfam\tenmib}
  \textfont\mibfam=\tenmib
  \scriptfont\mibfam=\eightmib
  \scriptscriptfont\mibfam=\sixmib
\def\frac#1#2{{\textstyle{#1 \over #2}}}
\def\bfJ{{\mib J}}
\def\bfB{{\mib B}}
\def\bfH{{\mib H}}
\def\bfI{{\mib I}}
\def\bfp{{\mib p}}
\def\bfk{{\mib k}}
\def\bigP{{\cal P}}
\def\pz{\partial}
\def\xhi{{\raise.35ex\hbox{$\chi$}}}
\def\pF{p_{\rm F}}
\def\kF{k_{\rm F}}
\def\nd{^{\vphantom{\dagger}}}
\def\yd{^\dagger}
\def\undertext#1{$\underline{\hbox{#1}}$}
\def\Tr{\mathop{\rm Tr}}
\def\ket#1{{\,|\,#1\,\rangle\,}}
\def\bra#1{{\,\langle\,#1\,|\,}}
\def\braket#1#2{{\,\langle\,#1\,|\,#2\,\rangle\,}}
\def\expect#1#2#3{{\,\langle\,#1\,|\,#2\,|\,#3\,\rangle\,}}
\gdef\journal#1, #2, #3, 1#4#5#6{               
    {\sl #1~}{\bf #2}, #3 (1#4#5#6)}            
\def\pra{\journal Phys. Rev. A, }
\def\prb{\journal Phys. Rev. B, }
\def\prl{\journal Phys. Rev. Lett., }
\def\cH{{\cal H}}
\def\cHt{\widetilde{\cal H}}
\def\wt{\widetilde}
\def\psit{\widetilde\psi}
\def\xhat{{\hat{\mib x}}}
\def\yhat{{\hat{\mib y}}}
\def\zhat{{\hat{\mib z}}}
\def\nhat{{\hat{\mib n}}}
\def\ie{{\it i.e.\/}}
\def\eg{{\it e.g.\/}}
\def\vth{\vartheta}
\def\half{\frac{1}{2}}
\def\oort{\frac{1}{\sqrt{2}}}
\def\thalf{\frac{3}{2}}
\def\fourth{\frac{1}{4}}
\def\sout{\nd_{\rm out}}
\def\sinn{\nd_{\rm in}}

\twocolumn[\hsize\textwidth\columnwidth\hsize\csname  
@twocolumnfalse\endcsname
\title{Non-Abelian Geometric Phases and Conductance of Spin-$\thalf$ Holes}

\author{Daniel P. Arovas}
\address{Department of Physics, University of California at San Diego,
 La Jolla CA 92093}

\author{Yuli Lyanda-Geller}
\address{Department of Physics, Loomis Physics Laboratory\\and\\
Beckman Institute for Advanced Science and Technology\\
University of Illinois at Urbana-Champaign, Urbana, IL 61801}

\date{\today}

\maketitle

\begin{abstract}
Angular momentum $J=\frac{3}{2}$ holes in semiconductor heterostructures
are showed to accumulate nonabelian geometric phases as a consequence of
their motion.  We provide a general framework for analyzing such a system
and compute conductance oscillations for a simple ring geometry.  We also
analyze a figure-8 geometry which captures intrinsically nonabelian
interference effects.
\end{abstract}

\vskip2pc]

\narrowtext

{\it Introduction} -- A nondegenerate quantum state undergoing adiabatic
evolution accumulates both a dynamical as well as a geometric (Berry's) phase
\cite{Berry,Wilczek}.  The geometric phase is responsible for a wide array of
interference phenomena, and has been measured in optics \cite{Tomita},
with neutral beams \cite{Bitter}, and by magnetic resonance \cite{Pines1,tycko}.
Most theoretical and experimental work has focussed on the adiabatic
evolution of nondegenerate eigenstates, where the geometric phase may be
interpreted as arising from a U(1) gauge potential.  The canonical example is
that of a spin in a constant amplitude magnetic field ${\mib B}(t)=B\nhat(t)$
whose direction $\nhat(t)$ varies in a closed path over the unit sphere
\cite{Berry}.  The phase is determined by the solid angle subtended by
$\nhat(t)$ in the course of its evolution.  

In certain high symmetry situations, an entire $n$-fold degenerate set of levels
may adiabatically evolve.  Wilczek and Zee \cite{WZ} showed that in such cases
the U(1) geometric phase generalizes to a U($n$) matrix,
\begin{equation}
U={\bigP}\exp\left(-i\oint A^i\,d\lambda_i\right)
\label{wilson}
\end{equation}
where $A^i_{\alpha\beta}=-i\expect{\alpha}{{\pz\over\pz\lambda_i}}{\beta}$
is the gauge potential matrix ($\ket{\alpha}$, $\ket{\beta}$ are adiabatic
eigenstates), $\{\lambda_i(t)\}$ is a set of slowly evolving parameters, and
$\bigP$ is the path ordering operator.  Such a system may exhibit nonabelian
effects in which {\it e.g.\/} one member of a multiplet evolves into another
upon completion of a cycle in parameter space.  One example of this phenomenon
is in crystalline nuclear quadrupole resonance (NQR), since the quadrupole
Hamiltonian $\cH=\half Q_{ij}I_i I_j$ is quadratic in the spin.  When the
electric field gradient tensor has cylindrical symmetry, the Hamiltonian can
be taken to be $\cH=\hbar\omega_Q(\nhat\cdot\bfI)^2$, where $\nhat$ lies in the
direction of the principal axis of $Q_{ij}$.  The nonabelian gauge structure
for this problem was discussed by Zee \cite{zee} and measured in the $I=\thalf$
nucleus ${}^{35}$Cl by Zwanziger, Koenig, and Pines \cite{zkp}.  Paths in which
$\nhat(t)$ rotates about more than one axis are essential if intrinsically
nonabelian aspects are to be captured.

In this paper we consider an alternative setting for an observation of
the nonabelian geometric phase.  We study angular momentum-$\thalf$ holes
confined to conducting loops embedded in a two-dimensional hole gas of a 
heterostructure.  A hole's momentum $\bfp(t)$ (or coordinate $\phi(t)$ in a
ring) acts as an adiabatically changing quantization axis for its angular
momentum $\bfJ$.  For motion around a ring, however, this
amounts to a rotation about only one axis.  To exhibit the nonabelian
effects lurking here, we propose to effectively place the system in a rotating
frame by imposing a static magnetic field $\bfH$ in the plane of the ring
\cite{nonplanar}.  Intrinsically nonabelian interference effects are measurable
in the conductance oscillations of the figure-8 device discussed below (see
fig. 2).  Another notable feature is that unlike the case studied in refs.
\cite{zee,zkp}, both hole doublets manifest a nonabelian holonomy.

The coupling of $\bfp$ to $\bfJ$, which arises naturally within a
$\bfk\cdot\bfp$ treatment of conduction electron and valence hole states, 
is analogous to the spin-orbit interaction.  It is qualitatively different,
however, from the spin-orbit splitting of electron states 
\cite{dressel,Luttinger}.  For electrons in zincblende crystals, the spin states
are split because of the inversion asymmetry, which in a quantum well or
heterostructure leads to a linear coupling between spin and momentum
\cite{Altshuler}; another source of linear coupling is
the asymmetry of the quantum well or heterojunction itself \cite{Bychkov84}.
The electron's momentum then acts as an in-plane component
of the magnetic field, and as the electron moves around a ring its spin
quantization axis traverses the unit sphere at a colatitude
$\theta=\tan^{-1}(H_z/\lambda p_\phi)$,
where $H_z$ is the physical magnetic field
(oriented perpendicular to the plane of the ring), $p_\phi$ is the azimuthal
component of the electron's momentum, and $\lambda$ is a coupling constant
\cite{yuli93}.  Although the spin-orbit coupling is by nature a relativistic
effect, it is effectively enhanced in a crystalline environment, and the
fictitious in-plane component of
the field can be of considerable magnitude \cite{enhancement}, although the
splitting of $\uparrow$ and $\downarrow$ states is still much less than the
kinetic energy of the electrons.
 
This situation is quite different for holes in group IV or III-V semiconductors,
which are characterized by a $4\times 4$ matrix Luttinger Hamiltonian, acting
on states in the $\Gamma_8$ representation of double groups of $T_d$ or $O_h$
\cite{dressel,Luttinger,soh}.  The effective Hamiltonian for bulk holes 
contains a term $(\bfp\cdot\bfJ)^2$, which distinguishes between light
($J^z=\pm\thalf$) and heavy ($J^z=\pm\half$) hole branches.  This term is
large and is present in bulk centrosymmetric materials.  In contrast to the
electron case, characterized by the effective magnetic field, the hole
hamiltonian is thus characterized by the effective quadrupole tensor field.
This leads to nonabelian effects, which can be probed in the conductance of
the double loop device, discussed below.

{\it Hamiltonian and its gauge structure} -- The effective Hamiltonian for the
valence band is written in terms of a spin-$\thalf$ operator
$\bfJ$ and the crystal momentum $\bfp$\cite{Luttinger,foot1}:
\begin{equation}
\cH=-(A+\frac{5}{4}B)\bfp^2 + {B}\,(\bfp\cdot\bfJ)^2\ ,
\label{lutt}
\end{equation}
where $A$ ($B$) is given by
$\frac{1}{4}\hbar^2(m_{\rm lh}^{-1}\pm m_{\rm hh}^{-1})$,
in terms of the light hole and heavy hole masses.
For simplicity, in this work we take the Hamiltonian for bulk holes in 
the spherical approximation; we also neglect all terms in the Hamiltonian
which arise due to the absence of inversion symmetry.
Consider now a geometry in which holes are confined to a ring lying
within the plane perpendicular to the $\zhat\parallel (001)$ axis.  The radial
coordinate $r$ is constrained to lie between $R$ and $R+a$, with $a\ll R$ and
the coordinate $z$ is also confined. Then the spin-dependent part of the
effective Hamiltonian reads
\begin{equation}
\cH\nd_{\rm eff}=\langle p_z^2\rangle B J_z^2+\fourth\langle p_r^2\rangle B 
(J^+ e^{-i\phi}+J^- e^{i\phi})^2
\label{Heff}
\end{equation}
valid to order $a/R$, with $\langle p_r^2\rangle\approx\pi^2\hbar^2/a^2$, and
$\langle p_z^2\rangle$ depending on the confining potential.  We now treat
the $\phi$ motion semiclassically and let $\phi(t)$ be a prescribed function
of time, with $d\phi/dt=\omega$ for motion around a ring \cite{kinetic}.
The spin Hamiltonian becomes $\cH=K(\nhat(t)\cdot\bfJ)^2+\Xi J_z^2$,
where $K=B\langle p_r^2\rangle$, $\Xi=B\langle p_z^2\rangle$, and
$\nhat(t)=\xhat\cos\phi(t) + \yhat\sin\phi(t)$ is the time-varying principle
quadrupole axis for our problem.  Now this quadrupole field rotates about
only one axis, and in order to extract nonabelian effects from this setting,
we must effectively introduce another axis of rotation by applying an
{\it in-plane} magnetic field $\bfH=H_x\xhat$, which adds a term
$\cH'=-g\mu_{\rm B} H_x J_x/\hbar$ to the Hamiltonian.  Note that there is
perpendicular component, hence no orbital effects of the magnetic field.
We then eliminate
$H_x$ by shifting to a rotating basis via the gauge transformation
$\ket{\psi}=\exp(-i\Omega t J_x/\hbar)\ket{\psit}$.  In this basis, the
Hamiltonian becomes
\begin{eqnarray}
\cHt(t)&=& V\yd(t)\,\cHt_0\, V(t)\nonumber\\
V(t)&=& \exp(i\phi(t) J_z/\hbar)\,\exp(i\Omega t J_x/\hbar)\nonumber\\
\cHt_0&=& K J_x^2 + \Xi J_z^2\ ,
\end{eqnarray}
with $\Omega=g\mu_{\rm B} H_x/\hbar$.
This Hamiltonian is similar to that explored in refs. \cite{tycko,zee,zkp}
in the context of $J=\frac{3}{2}$ nuclear quadrupole resonance.

We next compute the nonabelian gauge potential matrix $A_{\alpha\beta}(t)$:
\begin{equation}
A_{\alpha\beta}(\phi)=-i\expect{\wt{\alpha}(t)}{{d\over d\phi}}
{\wt{\beta}(t)}
\end{equation}
where $\ket{\wt{\alpha}(t)}=V\yd(t)\ket{\alpha}$ is an adiabatic eigenstate of
$\cHt(t)$.  The eigenstates of $\cHt_0$, expressed in eigenstates of $J_z$,
form two degenerate blocks ($\sigma=\pm$):
\begin{eqnarray}
\ket{1,\sigma}&=&u_\sigma\ket{-\frac{3}{2}}+v_\sigma\ket{+\half}
\nonumber\\
\ket{2,\sigma}&=&u_\sigma\ket{+\frac{3}{2}}+v_\sigma\ket{-\half}\nonumber\\
E_\sigma&=&\frac{5}{4}(K+\Xi)+\sigma\sqrt{K^2+\Xi^2-K\Xi}
\label{basis}
\end{eqnarray}
with $u_+=v_-=\cos\half\vth$, $v_+=-u_-=\sin\half\vth$, and
where $\tan\vth=\sqrt{3}\,K/(2\Xi-K)$.  The $4\times 4$ gauge potential matrix
$A_{\alpha\beta}$ is block diagonal in this basis with $2\times 2$ subblocks
\begin{eqnarray}
A^\pm_\omega(\phi)&=&\pmatrix{a&b\cr b^*&-a\cr}\nonumber\\
a&=&(\half\pm\cos\vth)\nonumber\\
b&=&\mp\frac{\sqrt{3}}{2}\,\frac{\Omega}{\omega}\,\sin\vth\, e^{-i\phi}-
\half\,\frac{\Omega}{\omega}\,(1\mp\cos\vth)\,e^{+i\phi}
\end{eqnarray}
This gauge potential determines the adiabatic evolution of wavefunctions 
of holes.  Finally, the U(2) phase accrued by a state evolving according to
the Hamiltonian $\cH+\cH'$ is
\begin{eqnarray}
U(t_1,t_0)&=&e^{i\kF L}\,e^{-i\phi_1 J_z/\hbar}\Lambda\yd_\vth\,
W_\omega(\phi_1,\phi_0)\,\Lambda\nd_\vth\,e^{+i\phi_0 J_z/\hbar}\nonumber\\
W_\omega(\phi_1,\phi_0)&=&{\bigP}\,\exp\left(-i\int_{\phi_0}^{\phi_1}
\!\!\!\!d\phi\,A_\omega(\phi)\right)
\label{evolution}
\end{eqnarray}
where $\Lambda\nd_\vth$ transforms from the $J^z$ eigenbasis to the basis of
eq. (\ref{basis}), $\kF$ is the Fermi momentum of the holes, $L$ is the
distance traveled, and where the path ordering operator places earlier
{\it times} to the right.  Adiabaticity is satisfied provided
$\omega,\Omega\ll\sqrt{K^2+\Xi^2-K\Xi}$.

\begin{figure} [t]
\centering
\leavevmode
\epsfxsize=9cm
\epsfbox[18 144 592 718] {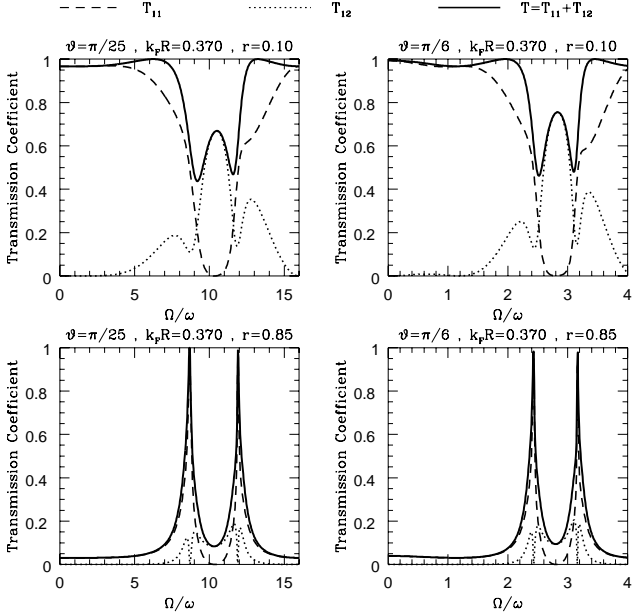}
\caption[]
{\label{fig1} Transmission coefficients $T_{11}$ and $T_{12}$ as a function
of $\Omega/\omega$ for a simple ring geometry.}
\end{figure}

{\it Conductance oscillations in a loop} -- We next consider the transport of
holes in the upper ($\sigma=+$) doublet through a ring confined to the
two-dimensional hole gas \cite{mil}. The ring is connected
to leads through two antipodally placed T-junctions, each described by the
$S$-matrix
\cite{bia} 
\begin{equation}
S=\pmatrix{r&\oort\sqrt{1-r^2}&\oort\sqrt{1-r^2}\cr
\oort\sqrt{1-r^2}&-\half(1+r)&\half(1-r)\cr
\oort\sqrt{1-r^2}&\half(1-r)&-\half(1+r)\cr}
\label{Smat3}
\end{equation}
where $r$ is the reflection amplitude for a wave incident from the incoming
lead, and where, for simplicity, we assume that $S$ is real and is diagonal
in the basis of Eq. \ref{basis}.  In Fig. \ref{fig1} we plot the transmission
probabilities $T_{\sigma\sigma'}=T_{\sigma'\sigma}$ -- due to the nonabelian
geometric phase an incoming hole in state $\ket{1+}$ may be transformed with
probability $T_{12}$ to the state $\ket{2+}$.  The conductance of the device
is given by $G={e^2\over h}\sum_{\sigma\sigma'}T_{\sigma\sigma'}=
{2e^2\over h} T$ (since both degenerate levels are occupied in the incoming
lead). In our computations we assumed that holes are confined to a plane
thickness $100$ \AA, Fermi energy (for holes) 2 meV, ring radius
$1\,\mu{\rm m}$, and the width of the ring and leads is $400$ \AA.  
This corresponds to a rotation frequency $\omega\approx 10^{10}$ Hz.

The $T_{\sigma\sigma'}$ are plotted for $\kF R=0.370$ (mod $\pi$) -- the
qualitative results are roughly insensitive to this parameter -- as a function
of $\Omega/\omega$ for two values of $\vth$ at both weak ($r=0.10$) and strong
($r=0.85$) coupling of leads to ring.  A ratio of $\Omega/\omega=4$ 
corresponds to a field of roughly 1 Tesla.  The resonances arise due to
interference effects both abelian and nonabelian in origin -- the intrinsically
nonabelian effects manifested in $T_{12}$ are not possible to isolate in this
geometry.

\begin{figure} [h]
\centering
\leavevmode
\epsfxsize=5cm
\epsfysize=5cm
\epsfbox[18 144 592 718] {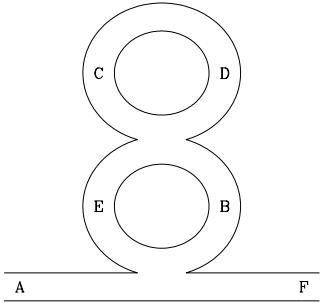}
\caption[]
{\label{fig2} Figure-8 device.}
\end{figure}

{\it Figure-8 device} -- The device depicted in Fig. \ref{fig2} probes
nonabelian interference effects.  Holes incident from the lead A may scatter
into branch B of the figure-8 or else continue on to lead F.  Neglecting
the effect of the magnetic field on the contacts themselves, we
assume a time reversal invariant (\ie\ symmetric) $S$-matrix for the ABEF
vertex of the form \cite{notgen}
\begin{equation}
\pmatrix{A\sout\cr B\sout\cr E\sout\cr F\sout\cr}=
\pmatrix{\vphantom{A\sout}u&v&0&t\cr \vphantom{A\sout}v&-u&t&0\cr
\vphantom{A\sout}0&t&u&-v\cr \vphantom{A\sout}t&0&-v&-u\cr}
\pmatrix{A\sinn\cr B\sinn\cr E\sinn\cr F\sinn\cr}
\label{Smat4}
\end{equation}
where $A\sinn$ is a two-component vector describing the incident flux
of holes in the upper doublet. The real parameters $t,u,v$ satisfy
$t^2+u^2+v^2=1$ and are assumed to be the same for both states of the
doublet.  $t\approx 1$ corresponds to weak coupling between the leads and the
figure-8, and $S_{1,3}=S_{2,4}=0$ means that there is negligible
backscattering between E and A and between B and F.  The vertex at
the center of the figure-8 is assumed to pass B to C and D to E with negligible
scattering into other channels, \ie\ the EDCB $S$-matrix corresponds to that
of eq. (\ref{Smat4}) with $t=1$.  This ensures that holes which enter
the figure-8 through branch B will execute a BDCE circuit before
entering the lead F or being re-scattered into B.  Such a scattering matrix 
for the BECD contact is realized when this contact is collimating, \ie\ it
conserves the momentum of holes.  The conservation of momentum in the course
of transmission through the contact holds if dimensions of the contact are
larger than the wavelength of holes, so that diffraction effects are
suppressed.  Such contacts are technologically feasible and were studied
in electron transport (for a review see \cite{Webb}). 
Under these conditions, we may write the relation between B and E as
\begin{eqnarray}
E\sinn&=&e^{i\kF L}\,W_\omega(0,-\pi)\,W_{-\omega}(-\pi,\pi)\,
W_\omega(\pi,0)\,B\sout\label{EBrel}\\
E\sout&=&e^{-i\kF L}\,W_{-\omega}(0,\pi)\,W_\omega(\pi,-\pi)\,
W_{-\omega}(-\pi,0)\,B\sinn\ .\nonumber
\end{eqnarray}

\begin{figure} [t]
\centering
\leavevmode
\epsfxsize=9cm
\epsfbox[18 144 592 718] {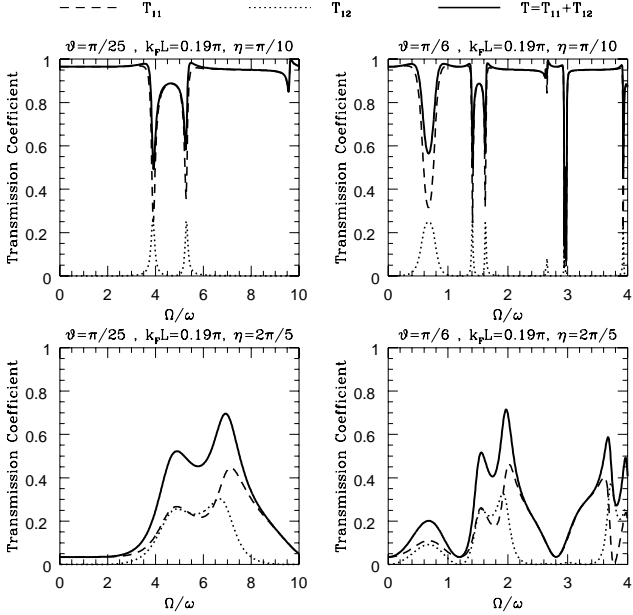}
\caption[]
{\label{fig3} Transport through the figure-8. The parameters of eq. 
(\ref{Smat4}) are related to $\eta$ by $t=\cos\eta$,
$u=v=\frac{1}{\sqrt{2}}\sin\eta$.}
\end{figure}

This result, in conjunction with eq. (\ref{Smat4}) determines the conductance
of the figure-8 device.  It is easy to see that when $\Omega=0$ (no in-plane
field), the gauge potential $A_\omega$ is diagonal and there are no nonabelian
effects -- the ``quadrupole field'' rotates only about the $\zhat$ axis.  This
reduces the products of $W$-matrices in eq. (\ref{EBrel}) to the unit matrix,
so that the only interference between the paths AF and ABCDEF is due to the
difference in their lengths.  In Fig. \ref{fig3}, resonances in the
transmission probability for the figure-8 device are shown (we use parameters
identical to those for the ring).  We find pronounced oscillations with the
variation of magnetic field.  These oscillations are a nonabelian effect, 
because the path executed by the holes corresponds to a zero net solid angle
subtended by the effective quadrupole field \cite{comment}, meaning that
abelian effects are cancelled.

We note that the interaction of holes moving in constrictions with localized
holes or nuclei via spin-spin interactions leads to possibilities of observing
nonabelian phases in nuclear (spin) resonance experiments.  Yet another option
is to study optical transitions of constricted holes.

We are grateful to Alex Pines for interesting discussions of non-abelian
settings and to Richard Webb for very useful remarks about adiabatic contacts.
We also thank P. Goldbart, A. Leggett, A. Manohar and L. J. Sham for
discussions.  YLG was supported by NSF under grants DMR91-57018 and
DMR94-24511.


\begin{references}
\bibitem{nonplanar} In the case of a non-planar conductor, non-abelian effects 
arise without the imposition of a magnetic field.  However, the requirement of
high mobility of holes rules out any such configuration at present.
\bibitem{enhancement} Compared to the relativistic case, one can estimate the
crystalline enhancement of the spin-orbit interaction by substituting the
effective band mass $m^*$ for the bare electron mass $m$, and the semiconducting
band gap $E_{\rm g}$ for the vacuum energy $2mc^2$,
resulting in an enhancement of several orders of magnitude.
\bibitem{soh} The $J=\half$ split-off hole states are not considered.
\bibitem{foot1} We will refer to the $J^z$ quantum number as ``spin''.
\bibitem{kinetic} For simplicity, we will analyze the problem assuming
that $B < A$. Then actually the same frequency $\omega$ characterises
semiclassical motion for both hole doublets of states.
\bibitem{notgen} The most general $N\times N$ $S$-matrix is specified by
$N^2$ real parameters.  Time reversal invariance implies $S=S^{\rm t}$;
this restriction reduces the dimension of the space to $\half N(N+1)$.
Thus, the $S$-matrices of eqs. (\ref{Smat3},\ref{Smat4}) are nongeneric.
\end{references}
\end{document}